# Dynamics of antibody binding and neutralization during viral infection

**Zhenying Chen[1#], Hasan Ahmed[1#], Cora Hirst[1], Rustom Antia[1*]**

Department of Biology, Emory University, Atlanta, GA, USA

# contributed equally

* rantia@emory.edu

## Abstract

*In vivo* in infection, virions are constantly produced and die rapidly. In contrast, most antibody binding assays do not include such features. Motivated by this, we considered virions with $n=100$ binding sites in simple mathematical models with and without the production of virions. In the absence of viral production, at steady state, the distribution of virions by the number of sites bound is given by a binomial distribution, with the proportion being a simple function of antibody affinity ($K_{on}/K_{off}$) and concentration; this generalizes to a multinomial distribution in the case of two or more kinds of antibodies. In the presence of viral production, the role of affinity is replaced by an infection analog of affinity (IAA), with $IAA=K_{on}/(K_{off}+d_v+r)$, where $d_v$ is the virus decaying rate and $r$ is the infection growth rate. Because *in vivo* $d_v$ can be large, the amount of binding as well as the effect of $K_{off}$ on binding are substantially reduced. When neutralization is added, the effect of $K_{off}$ is similarly small which may help explain the relatively high $K_{off}$ reported for many antibodies. We next show that the n+2-dimensional model used for neutralization can be simplified to a 2-dimensional model. This provides some justification for the simple models that have been used in practice. A corollary of our results is that an unexpectedly large effect of $K_{off}$ *in vivo* may point to mechanisms of neutralization beyond stoichiometry. Our results suggest reporting $K_{on}$ and $K_{off}$ separately, rather than focusing on affinity, until the situation is better resolved both experimentally and theoretically.

# 1 Introduction

Antibodies are believed to play a pivotal role in controlling and preventing infection and moreover are relatively easy to measure, compared to other components of the adaptive immune system such as B and T cells. For these reasons, extensive research, experimental as well as computational and theoretical research, has involved antibody responses to a variety of pathogens including viruses.

Experimental *in vitro* assays for antibody binding and neutralization often involve incubating virus for 30 minutes to 2 hours, or even overnight, before binding or neutralization accessed (Russell et al. 1967; Kaufmann et al. 2017; Cuevas et al. 2022; Alhajj et al. 2023), allowing the dynamics of antibody binding to approach an equilibrium. However, *in vivo* in infections, virus production and death occur while antibodies bind and neutralize the virions. For this reason, in the first part of the paper, we use mathematical models to assess the differences between antibody activity at binding equilibrium in the absence of virus production versus antibody activity in an infection in which virions are constantly being produced and die.

Mathematical models of viral infection often use highly simplified models of antibody binding and function (Perelson 2002; Ciupe et al. 2011; Schwartz et al. 2015; Best and Perelson 2018; Guo et al. 2020; Meadows and Schwartz 2022) , such as a single mass action term. In contrast, a number of papers, especially mathematical models for HIV, have used more complicated 'stoichiometric' model (Magnus and Regoes 2010; Ciupe et al. 2014; Webb et al. 2015; Brandenberg et al. 2015, 2017)in which the number of bound sites on a virion is related to the virion's infectivity. Simple models have several advantages, such as being more conducive to exact analytical solutions as well as improved interpretability and quantitative estimation. On the other hand, they can also introduce results that are inconsistent with or even the opposite of models that incorporate known biological complexities (Nikas et al. 2023). For this reason, in the second half of the paper, we consider the extent to which the stoichiometric model used in the first half of the paper, parameterized using the data from Pierson et al. (Pierson et al. 2007), can be approximated using very simple models.

## 2 Results

### 2.1 Modeling antibody binding

Considering the fundamental dynamics of antibody binding, we have developed a staged ordinary differential equation (ODE) model. In this model, each virion stage represents the number of sites bound by antibodies (refer to Figure 1 for a schematic and Table 1 for values and units of each parameter). $V_0$ designates the compartment of newly born virions with no antibodies bound, while $n$ is the total number of bind sites per virion. Infected cell $I$ continuously produces new virions ($V_0$) at a rate of $\rho$. Intermediate stages between $V_0$ and $V_n$, denoted as $V_i$, can progress to the next stage $V_{i+1}$ when an antibody encounters the virion and binds to one of the free sites at a rate $K_{on}A$, where $A$ is the antibody concentration. Conversely, $V_{i+1}$ can revert to the previous stage $V_i$ if the antibody dissociates from a bound site at a rate $K_{off}$. We assume a lack of steric interference and that all virions are decaying at a rate of $d_v$.

$$\frac{dV_0}{dt} = \rho I - nK_{on}AV_0 + K_{off}V_1 - d_vV_0$$
$$\frac{dV_i}{dt} = \big(n-(i-1)\big)K_{on}AV_{i-1} - (n-i)K_{on}AV_i + (i+1)K_{off}V_{i+1} - iK_{off}V_i - d_vV_i, 0 < i < n$$
$$\frac{dV_n}{dt} = K_{on}AV_{n-1} - nK_{off}V_n - d_vV_n \qquad (Model\ 1)$$

For *in vivo* infection, we assume $I(t) = I(0)e^{rt}$, otherwise $\rho = 0$.

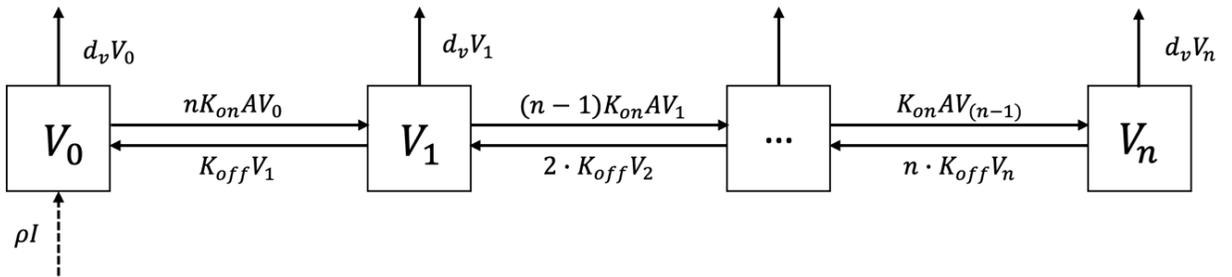

**Figure 1. Schematic of the model depicting the basic dynamics of antibody binding (Model 1).** Antibodies bind to free sites at a rate $K_{on}A$ and dissociate at a rate $K_{off}$. $n$ signifies the total number of available binding sites on the virion for antibodies. Virions at all stages are decaying at rate $d_v$ and new virions are reproduced at rate $\rho I$.

**Table 1. Parameters Used for Simulating Model 1**

| Parameters | Description | Value | Unit | Reference |
|---|---|---|---|---|
| $I(0)$ | Infected cells | NA or 1 | CU | / |
| $V_0(0)$ | Free virus | 100 | VU | / |
| $K_{on}$ | antibody on rate | 1 | AU$^{-1}$ day$^{-1}$ | / |
| $K_{off}$ | antibody dissociation rate | 12 | day$^{-1}$ | (Schieffelin et al. 2010) |
| $A$ | free antibody concentration | varies | AU | / |
| $n$ | number of binding sites on a virion | 100 | sites virion$^{-1}$ | (Zhang et al. 2003, 2004) |
| $d_v$ | virus death rate | 10 | day$^{-1}$ | (Banerjee et al. 2016) |
| $\rho$ | virus production rate | 0 or 800 | VU CU$^{-1}$ day$^{-1}$ | / |
| $r$ | overall infection growth rate | NA or 2.5 | ln day$^{-1}$ | |

When $\rho$ is equal to 0, values of $I(0)$ and $r$ are not relevant to the simulation.

## 2.2 The number of bound sites on a virion follows a binomial or multinomial distribution at $\rho$=0 equilibrium

In many antibody titer assays, the process of antibody binding to virions approaches equilibrium prior to measurement. Given the absence of virus production in this process, we assume the virus reproduction rate ($\rho$) is equal to 0. When the dynamics of virus and antibody binding reach equilibrium, a distribution of virion stages should emerge. Although directly solving the staged ODE model for this distribution is challenging, a simple solution arises by considering the proportion of sites bound on a virion ($p$) across all virions. In the context of a single type of antibody, $\frac{dp}{dt} = K_{on}A(1-p) - K_{off}p$. At steady state, the proportion of bound sites is equal to $\frac{K_{on}A}{K_{on}A+K_{off}}$. As each site on a virion can exist in one of two mutually exclusive states - bound or unbound by an antibody - the distribution of virion stages follows the binomial distribution:

$$binomial\ (N = n, p = \frac{K_{on}A}{K_{on}A+K_{off}}) \hspace{2cm} \text{(Eq. 1)}$$

See Figure 2 for a binomial distribution example.

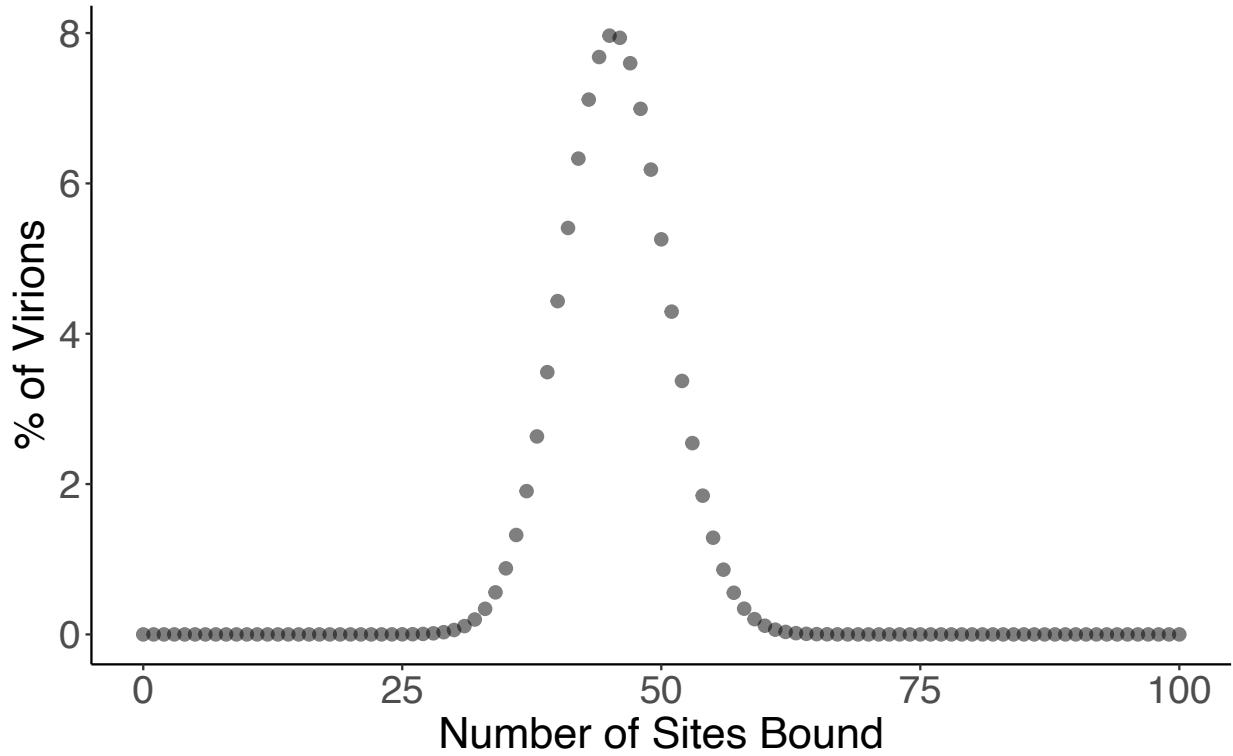

**Figure 2. Distribution of virion stages in the presence of a single type of antibody follows a binomial distribution.** The black dots represent the binomial distribution of virion stages ($N = n, p = \frac{K_{on}A}{K_{on}A + K_{off}}$), with an antibody concentration of 10 AU, $K_{on}$ at 1 AU$^{-1}$ day$^{-1}$, and $K_{off}$ at 12 day$^{-1}$.

This result generalizes to scenarios where multiple types of antibodies with potentially different $K_{on}$ and $K_{off}$ compete for binding to the same virion site or where sites are in close proximity such that they affectively act as a same site. In this case, the distribution of bound sites by each type of antibody follows a multinomial distribution:

$$Multinomial\left(N = n, \; p_i = \frac{X_i A_i}{1 + \sum_{j=1}^{m} X_j A_j}\right), \; X_i = \frac{K_{on_i}}{K_{off_i}} \qquad \text{(Eq. 2)}$$

, where *m* is the number of types of antibodies. See supplement section 1.2 for derivation and Figure 3 for a multinomial distribution example.

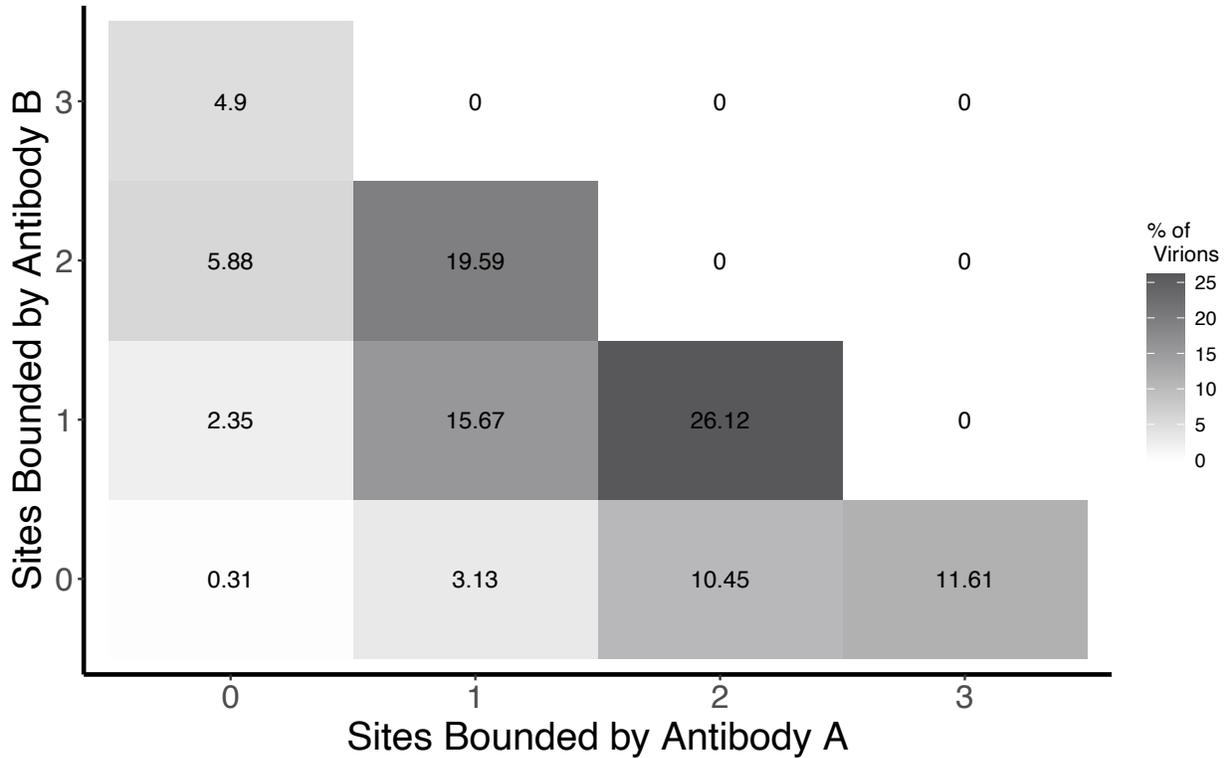

**Figure 3. Distribution of virion stages in the presence of two types of antibodies follows a multinomial distribution.** Steady state distribution for a scenario involving two types of antibodies, designated as Antibody A and Antibody B. Both Antibody A and Antibody B are present at the same concentration of 20 AU, but their antibody affinities are distinct. For Antibody A, $K_{on}$ was set to 2 AU$^{-1}$ day$^{-1}$, and $K_{off}$ was set to 12 day$^{-1}$. Meanwhile, Antibody B had $K_{on}$ set to 1 AU$^{-1}$ day$^{-1}$, and $K_{off}$ set to 8 day$^{-1}$. The results are visualized on a heatmap. In the heatmap, each cell's color and value indicate the percentage of virions with sites bound by specific amounts of Antibody A and Antibody B.

## 2.3 Analytical solution for proportion of bound sites during infection

During an infection, virions constantly produced by infected cells with newly produced virions assumed to be initially free of antibody. Given the relatively slow and delayed increase in antibody titers relative to virus load (Edupuganti et al. 2013), we model the antibody concentration as constant. In this case the distribution across virion stages approaches a distribution that is no longer binomial and is instead left skewed relative to a binomial distribution (Figure 4). Nonetheless, determining the mean of the distribution is feasible. In order to do so, we considered the proportion of sites bound on virions of a specific virion age $a$ as $p(a) = \int_0^a K_{on}A(1 - p(t)) - K_{off} \, dt$. We then considered the age distribution, which is a probability density function that depends on the growth rate of infected cells ($r$) and the decay rate of the virus ($d_v$), as $f(a) = (r + d_v)e^{-ra}e^{-d_v a}$. Now knowing these two quantities, we calculated the average proportion of bound sites $p^*$ (refer to the supplement section 2.1 for detailed calculation):

$$p^* = \int_0^\infty p(a) \cdot f(a) \, da = \frac{K_{on}A}{K_{on}A + K_{off} + r + d_v} \qquad \text{(Eq. 3)}$$

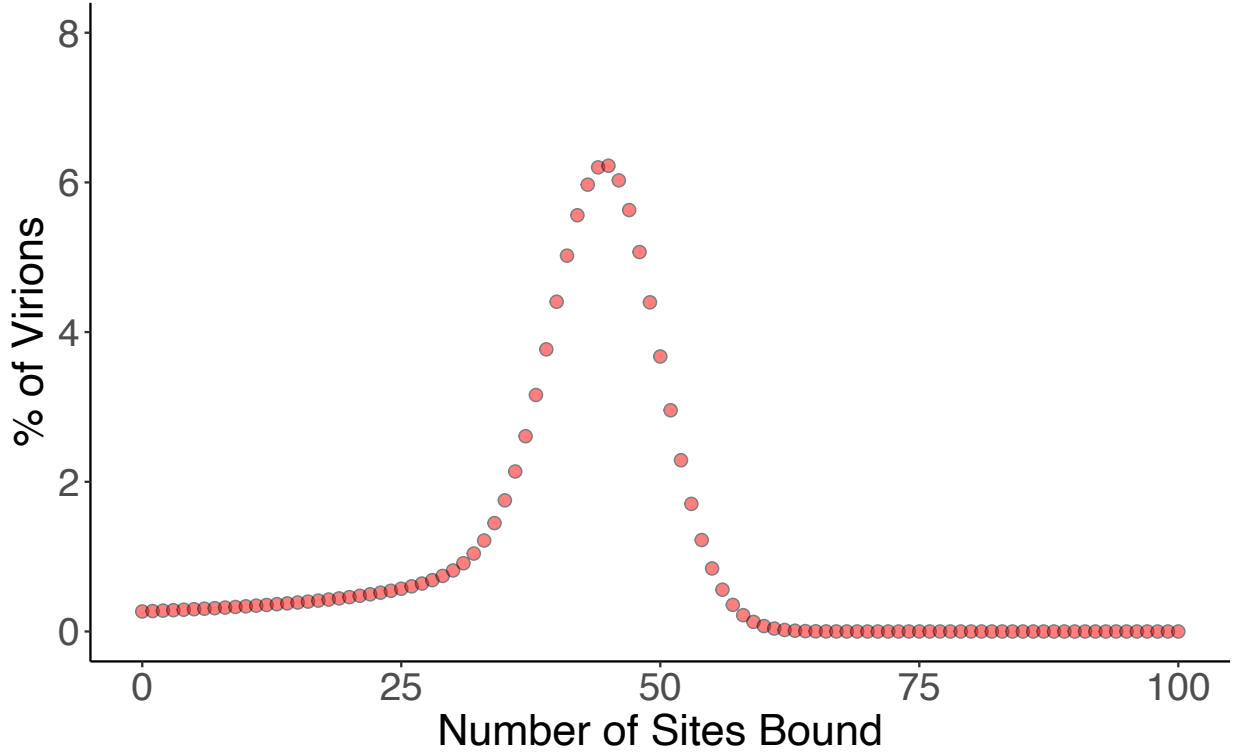

**Figure 4. During infection the distribution of virions deviates from a binomial distribution.** Simulations have been conducted using Model 1 when there is virus infection. The virus reproduction rate is set to 800 virions/day, and we assume that infected cells are increasing exponentially at rate r = 2.5 ln day$^{-1}$. The antibody concentration is set to 10 AU, with an antibody on rate $K_{on}$ equals to 1 AU$^{-1}$ day$^{-1}$ and $K_{off}$ equals to 12 day$^{-1}$.

## 2.4 Infection analog of affinity

In the absence of viral production, the mean proportion of sites bound is a function of antibody concentration and antibody affinity (Eq.1). In the presence of virial production affinity is replaced by a quantity we refer to as the infection analog of affinity (IAA):

$$\frac{K_{on}}{K_{off} + r + d_v} \qquad \text{(Eq. 4)}$$

This analogy is especially clear if we consider, instead of the proportion of sites bound, the odds of a site being bound in the absence of viral production the odds is $\frac{p}{1-p} = \frac{\frac{K_{on}A}{K_{on}A + K_{off}}}{1 - \frac{K_{on}A}{K_{on}A + K_{off}}} = \frac{K_{on}}{K_{off}} A$, whereas in an infection it is $\frac{p^*}{1-p^*} = \frac{\frac{K_{on}A}{K_{on}A + K_{off} + r + d_v}}{1 - \frac{K_{on}A}{K_{on}A + K_{off} + r + d_v}} = \frac{K_{on}}{K_{off} + r + d_v} A$. Hence in both cases, the odds is equal to the antibody concentration multiplied by either affinity of its infection analog. This dynamic analog of affinity is not only by determined by the ratio between $K_{on}$ and $K_{off}$, but also limited by virus growth rate $r$ as well as the virus decay rate $d_v$.

Realistically, the virus growth rate $r$ is a function of $K_{on}$, $K_{off}$, $d_v$ and $A$, and we expect the value of $r$ to be relatively small compares to the sum of $K_{off}$ and $d_v$ ($r < 3.5$/day (Moore et al. 2018; Kissler et al. 2021)), additionally $r = 0$ is a special value that corresponds to sterilizing levels of immunity, for these reasons we eliminated the parameter of $r$ in the dynamic analog of affinity and further simplified it as

$$\frac{K_{on}}{K_{off}+d_v} \qquad \text{(Eq. 5)}$$

, which we refer to as the simplified infection analog of affinity (IAA2). In other words, in the scenario of infection, the total amount of bound antibodies on a virion is not only reduced by $K_{off}$ but also by $d_v$.

Notably this result again generalizes to the situation where there are multiple types of antibodies (potentially different $K_{on}$ and $K_{off}$) competing for the same binding site. In this case the proportion of sites bound by antibody $i$, when virus is being actively produced, is:

$$p_i^* = \frac{Y_i A_i}{1+\sum_{j=1}^{m} Y_j A_j}, \; Y_i = \frac{K_{on_i}}{K_{off_i}+d_v+r} \qquad \text{(Eq. 6)}$$

Once again, the infection analog of affinity takes the place of affinity and can be further simplified by setting $r = 0$. See supplement section 2.2 for the derivation.

## 2.5  Virus infectivity decreases in a convex manner as proportion of bound sites increases

Previous experiments sought to determine the stoichiometry of neutralization for West Nile Virus (Figure 5A) (Pierson et al. 2007). In this experiment, the envelope (E) protein on the reporter virus particle, capable of only one round of infection, was modified into two types: wild-type (neutralization-sensitive) and mutant type (neutralization-resistant). Virus particles with varying percentages of wild-type and mutant-type E proteins were produced using plasmids, and preincubated with monoclonal neutralizing antibodies for 2 hours. The virus-antibody mixture was then transferred to target cells, and relative infection was detected using flow cytometry (see Figure S4 for visualized schematic of the procedure). Assuming a high concentration of antibodies will bind to most wild-type E proteins, we approximated the infectivity of a virus with $i$ percent of sites bound by the infectivity of virus with $i$ percent wild type E proteins. In this manner, we derived the approximate $\beta(i)$ function, describing virus infectivity ($\beta$) for $i$ proportion of virion sites bound by neutralizing antibodies (Figure 5B) (supplement section 3). Interestingly, $\beta(i)$ is convex rather than sigmodal as might have been expected.

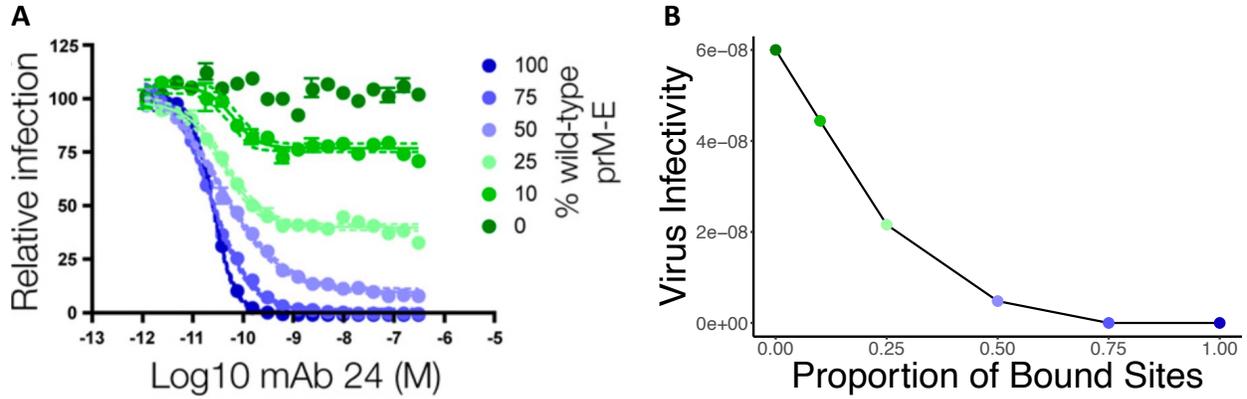

**Figure 5. Impact of changing the antibody-binding sites on the neutralization potency of monoclonal neutralizing antibody. (A)** The figure is taken from figure 5 panel A of the paper *"The stoichiometry of antibody-mediated neutralization and enhancement of West Nile virus infection"* by Pierson et al. (Pierson et al. 2007). Each line in the graph corresponds to virus particles comprising a blend of wild-type envelope (E) protein (neutralization sensitive) and mutant-type E protein (neutralization resistant), ranging from 0% wild-type E protein (dark green) to 100% wild-type E protein (dark blue). These lines show changes in relative infection of the virus particles following pre-incubation with a series of 2-fold antibody dilutions. **(B)** $\beta(i)$ function describes the decrease in virus infectivity as the proportion of bound sites increases. The function is derived through taking the end points of each line from panel A and replotted on a different scale. The color of each point corresponds to the original color that was used to show the percentage of wild-type E protein in panel A.

## 2.6 Modeling antibody neutralization

After deriving the $\beta(i)$ function, we can integrate this function into Model 1 to assess the antibody neutralization effect (refer to Figure 6 for schematics and Model 2 for equations). The virus's infectivity is now constrained by the concentration of antibodies: as the antibody concentration increases, the proportion of bound sites on a virion rises, resulting in lower virus infectivity. Instead of $I(t) = I(0)e^{rt}$, infected cells ($I$) now increase according to the number and infectivity of virions and die at rate $d_i$. We assume that target cell depletion is negligible, and hence the number of target cells ($U$) is a constant (Moore et al. 2020).

$$\frac{dV_0}{dt} = \rho I - nK_{on}AV_0 + K_{off}V_i - d_vV_0$$

$$\frac{dV_i}{dt} = \big(n-(i-1)\big)K_{on}AV_{i-1} - (n-i)K_{on}AV_i + (i+1)K_{off}V_{i+1} - iK_{off}V_i - d_vV_i, \ 0 < i < n$$

$$\frac{dV_n}{dt} = K_{on}AV_{n-1} - nK_{off}V_n - d_vV_n$$

$$\frac{dI}{dt} = \sum_{i=0}^{n} \beta(i)UV_i - d_i I \qquad\qquad (Model\ 2)$$

While Model 2 depicts a scenario more similar to an *in vivo* infection, the model for an *in vitro* neutralizing antibody assay should differ. In an *in vitro* assay, the binding of antibody and virus approaches equilibrium before virus infection. To account for this aspect in the model, we developed the "magic model" where newly produced virions instantly and "magically" reach equilibrium, forming a binomial distribution as described in Eq. 2 (Model 3).

$$\frac{dV}{dt} = \rho I - d_v V$$

$$V_i = \binom{n}{i}\left(\frac{K_{on}A}{K_{on}A + K_{off}}\right)^i \left(1 - \frac{K_{on}A}{K_{on}A + K_{off}}\right)^{n-i} \cdot V$$

$$\frac{dI}{dt} = \sum_{i=0}^{n} \beta(i)UV_i - d_i I \qquad\qquad (Model\ 3)$$

Using both the full model and the magic model, we examine parameter sensitivities regarding virus growth rate (Figure 7). In both models, the antibody concentration required for a 50% reduction in virus growth is much lower than half of the concentration needed for sterilizing immunity. Notably, in the full model, antibody dissociating rate $K_{off}$ has minimal effect in decreasing virus growth rate compared to antibody binding rate $K_{on}$. However, in the magic model, the effect of $K_{on}$ and $K_{off}$ in controlling virus growth rate is equivalent. Furthermore, the minimum antibody concentration needed for sterilizing immunity (SLA) is more than twice as high according to the full model than the magic model.

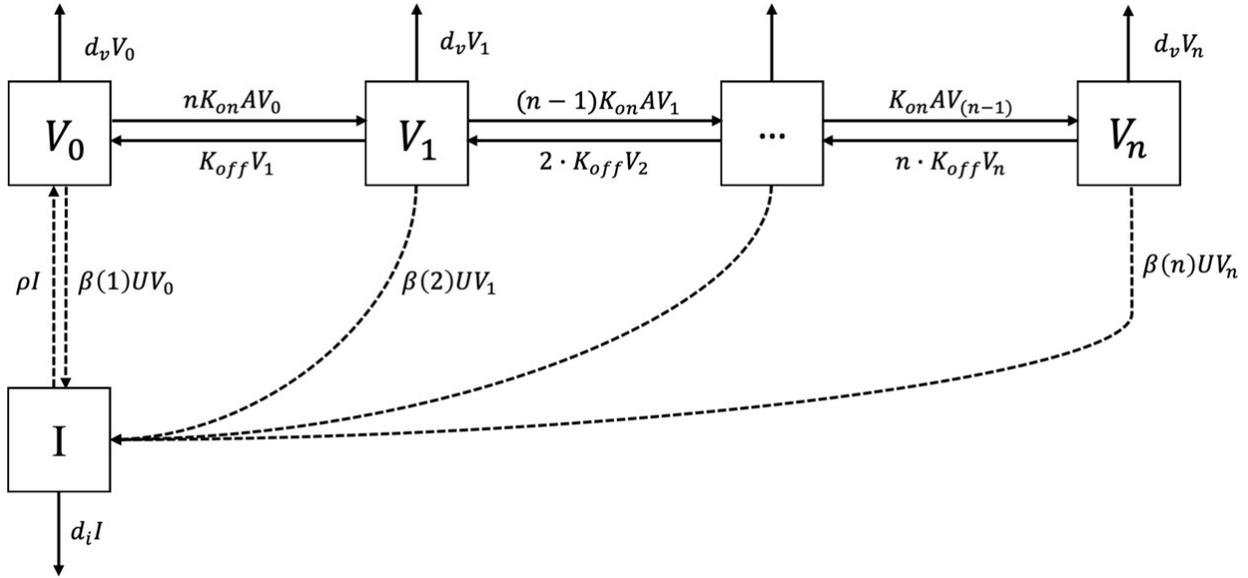

**Figure 6. Schematic of the full model depicting the basic dynamics of antibody neutralization (Model 2)**
Antibodies bind to free sites at a rate $K_{on}A$ and dissociate at a rate $K_{off}$. $n$ signifies the total number of available binding sites on the virion for antibodies. Virions at all stages are decaying at rate $d_v$ and new virions are reproduced at rate $\rho I$. Virus infectivity is considered as a function $\beta(i)$ where $i$ is the number of antibodies bound on a virion.

**Table 2. Parameters Used for Simulating Model 2 and Model 3**

| Parameters | Description | Value | Unit | Reference |
|---|---|---|---|---|
| $U$ | target cell population | $10^6$ | CU | / |
| $d_i$ | infected cell death rate | 1 | day$^{-1}$ | (Banerjee et al. 2016) |
| $\beta(0)$ | virus maximum infectivity rate | $6 \times 10^{-8}$ | VU$^{-1}$ day$^{-1}$ | * |

* Fitted to give a virus growth rate ($r$) equals to 2.5 $ln$/day.

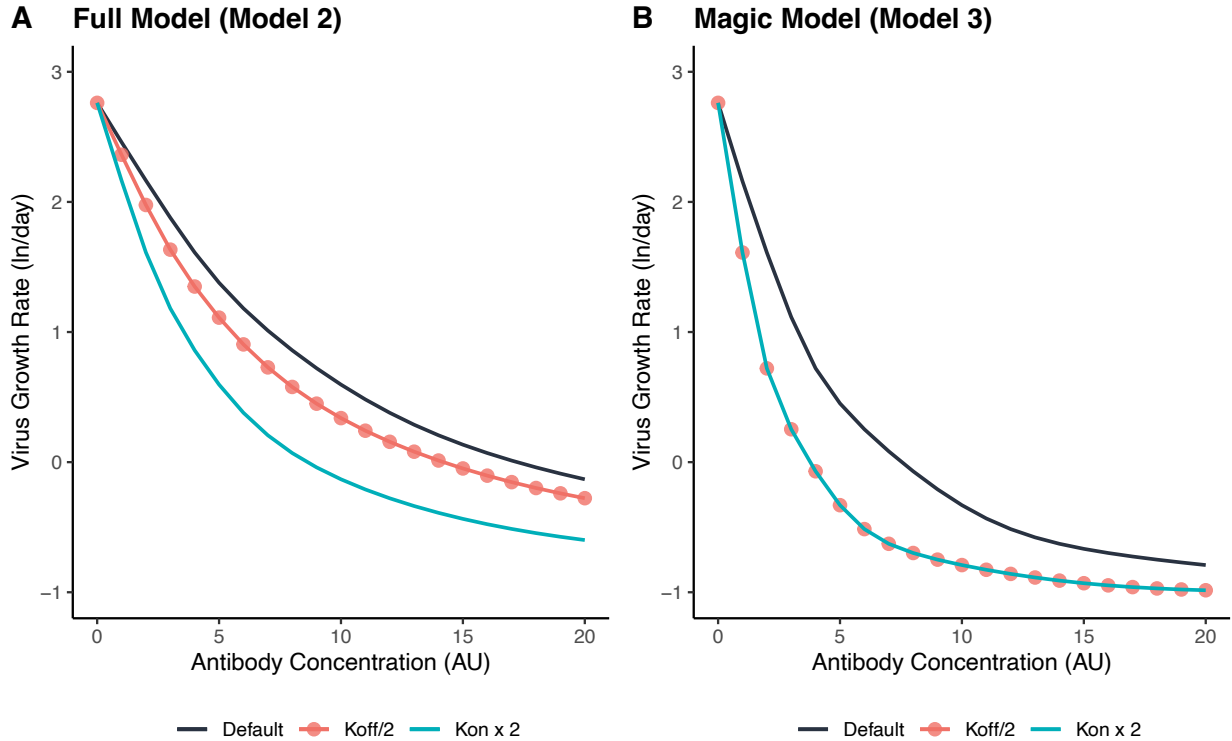

**Figure 7. Comparison between the simulations of magic model and full model as well as their corresponding parameter sensitivities.** Simulations show to the change in virus growth rate (in natural log scale) as antibody concentration increases using **(A)** the full model (Model 2) and **(B)** the magic model (Model 3). The black line represents the simulation conducted using default parameter values as listed in Tables 1 and 2. The red line represents simulations conducted with the parameter value of $K_{off}$ set to be half the default value. The green line represents simulations conducted with the parameter value of $K_{on}$ to be twice the default value.

## 2.7 Effect of $K_{off}$

As demonstrated previously, IAA2 ($\frac{K_{on}}{K_{off}+d_v}$) suggests that the impact of $K_{off}$ is relatively small in contributing to antibody binding, and the same pattern is seen for neutralization (Figure 7A). To further investigate this pattern, we use SLA$^{-1}$ as a measure of neutralization potency and vary $K_{off}$ from its default value of 12/day to 6/day and 24/day (Table 3). SLA$^{-1}$ in the magic model, like affinity, decreases 2-fold when $K_{off}$ is increased 2 fold and vice versa. In contrast SLA$^{-1}$ in the full model is even less sensitive than IAA2 to changes in $K_{off}$ which is attributable to the convexity of $\beta(i)$. When $\beta(i)$ decreases from 1 to 0 linearly SLA$^{-1}$ and IAA2 are equally sensitive whereas when it decreases concavely SLA$^{-1}$ is more sensitive (supplement section 4). Hence the relationship between IAA2 and SLA$^{-1}$ in the full model is roughly analogous to the relation between affinity and SLA$^{-1}$ in the magic model but complicated by the shape of the $\beta(i)$ function.

**Table 3: Affinity, IAA2, and SLA in the Full Model and the Magic Model as $K_{off}$ varies**

| Value of $K_{off}$ (day$^{-1}$) | SLA$^{-1}$ Calculated Using Full Model (Model 2) | SLA$^{-1}$ Calculated Using Magic Model (Model 3) | IAA2 ($\frac{K_{on}}{K_{off}+d_v}$) | Affinity ($\frac{K_{on}}{K_{off}}$) |
|---|---|---|---|---|
| 6 | 0.070 (+1.21)$^2$ | 0.26 (+2.0) | 0.0625 (+1.38) | 0.17 (+2.0) |
| 12 (default) | 0.058 | 0.13 | 0.045 | 0.083 |
| 24 | 0.042 (-1.39) | 0.066 (-2.0) | 0.029 (-1.55) | 0.042 (-2.0) |

SLA is the minimum antibody concentration for sterilizing immunity. IAA2 is the simplified infection analog of affinity. The numbers in the parenthesis are the fold change relative to the default value with + indicated fold increase and - fold decrease. The units for SLA$^{-1}$ and affinity are AU$^{-1}$. Note that in this context, IAA and IAA2 are equivalent as $r = 0$ at the SLA.

## 2.8 Simplified models can yield similar results of the full model

To simplify Model 2, we considered two strategies. First, we consider that the entire effect of antibody is given by a mass action term, $KAV$, which moves infectious virus into a non-infectious or cleared state; here $K$ is a constant for antibody potency, $A$ is antibody concentration, and $V$ is viral load. We refer to this model (Model 4) as the mass action model. Second, we consider replacing $\sum_{i=0}^{n} \beta(i)V_i$ in the full model with $f(A)V$ where $f(A)$ is a simple function of antibody concentration; in this case $f(a) = \frac{\beta}{1+CA}$ works well. We refer to this model (Model 5) as the mean field approximation model. See Figure 8 for schematics for both simplified models.

$$\frac{dI}{dt} = \beta UV - d_i I$$

$$\frac{dV}{dt} = \rho I - d_v V - KAV \quad (Model\ 4)$$

$$\frac{dI}{dt} = \frac{\beta}{1+CA} UV - d_i I$$

$$\frac{dV}{dt} = \rho I - d_v V \quad\quad (Model\ 5)$$

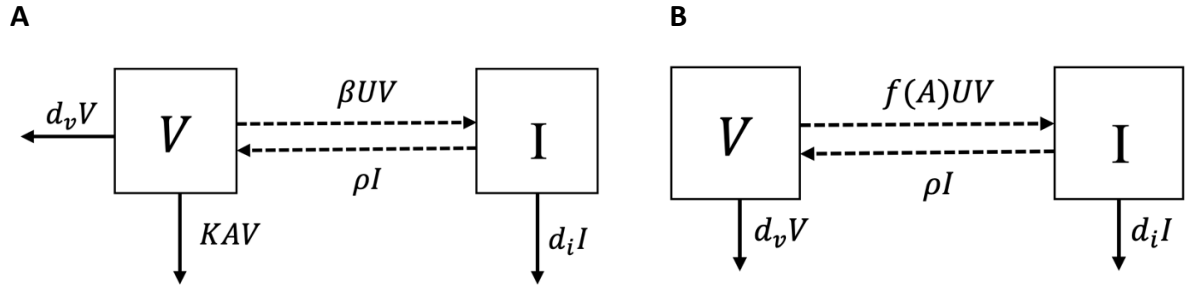

**Figure 8: Schematic of the mass action model (Model 4) and the mean field approximation model (Model 5).**
**(A)** The mass action model assumes that infectious virus *V* is neutralized at rate *KAV* and effectively removed from the system. The infectious virus decays at rate $d_v$. The infected cells produce free virus and decay at rate $d_i$. **(B)** The mean field approximation model assumes that antibody concentration decreases virus infectivity according to the function *f(A)*, while the infected cells produce virus. The virus and infected cells both decay at a constant rate of $d_v$ and $d_i$ respectively.

By fitting the sequester model and the reduce infectivity model with the full model, both simplified models offer a close approximation of the full model's results (Figure 9). However, it is also essential to acknowledge that the trade-off with these simplified models lies in their inability to explore the influence of antibody affinity and the proportion of bound sites on a virion in terms of reducing virus growth rates. Therefore, when selecting a model for future simulations or data fitting, the choice should be contingent upon the specific research question of interest.

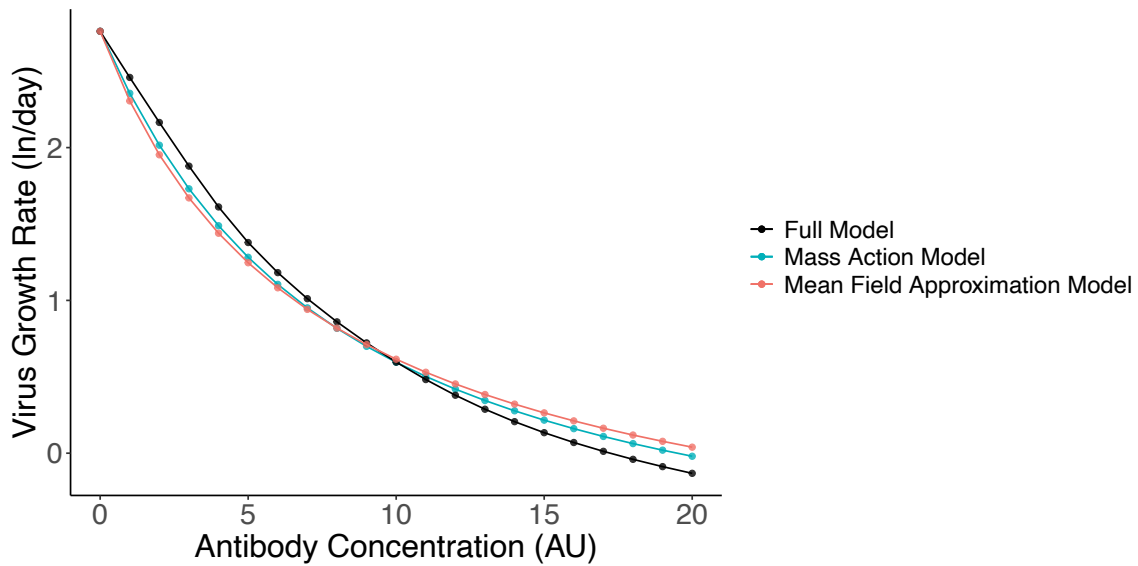

**Figure 9. Comparative simulation outcomes: the full model (Model 2), the mass action model (Model 4), and the mean field approximation model (Model 5).** Simulations regarding change in virus growth rate at various antibody concentrations have been examined using three models: the full model (green), the reduced infectivity model (red), and the sequester model (black). Antibody concentration is plotted linearly on the x-axis, while virus growth rate is plotted on a natural log scale along the y-axis. The constants in the simplified models were determined by minimizing the residuals of their simulations against the full model simulations. For the reduced infectivity model, the parameter value of C is equal to 0.18. For the sequester model, the parameter value of K is set to 1.95.

## 3  Discussion

In this paper, we address two questions. First, we consider how the production and death of virus alters the relationships that exist between affinity and antibody binding/neutralization. Second, we assess the ability of very simple models to capture the behavior of a more complex model of neutralization.

In the first part, we show an analogy between $\frac{K_{on}}{K_{off}+r+d_v}$ (IAA) and affinity. When virus production and death were allowed to disrupt the typical binding equilibrium, IAA took the place of affinity, in some regards approximately (e.g. Table 3) and in others exactly (e.g. Eq.2 vs. Eq.6). IAA implies that the effect of $K_{off}$ is relatively small compared to $K_{on}$. McKinley et al. reached a similar conclusion in a paper that assesses the effect of broadly neutralizing antibodies such as VRC01 on the probability of acquiring HIV infection (McKinley et al. 2014). It seems likely that such a conclusion will show up in many models that assume that neutralization is a function of the amount of antibody bound to particular sites. In our paper, we conservatively used an *in vivo* virus death rate of 10 day[-1] (half-life of 100 minutes) whereas Sukarai et al. reported a half-life of only 3 minutes for non-replicating adenovirus in mice (Sakurai et al. 2003). We speculate that the *in vivo* virus death rate is much greater than the *ex vivo* virus death rate possibly as a result of nonproductive viral entry into cells. A corollary of our results is that if the effect of $K_{off}$ is much bigger than suggested, the mechanism of neutralization may be something substantially different from the antibody occupancy model in which neutralization is primarily or solely a function of whether sites are covered or not (Burton 2023). Alternatively, if we view antibody as making certain processes less energetically favorable for the virus, low $K_{off}$ of antibody may be disproportionally important. Until the situation is better resolved, we recommend reporting $K_{on}$ and $K_{off}$ separately as opposed to focusing on the affinity ratio.

In the second part, we show that greatly simplified models of neutralization, the mass action and mean field approximation models, give similar virus growth dynamics as compared to our full model. It should be noted that our results here are based on experimental data from Pierson et al. and hence not particularly dependent on the antibody occupancy model (Pierson et al. 2007). Although our mass action and mean field approximation models cannot answer specific questions about the binding and neutralization processes such as the effect of $K_{off}$, these greatly simplified models can otherwise be used as plausible approximations.

## Acknowledgements
We acknowledged grants from the NIH: U01 AI150747 and U01 AI144616.

## Data Availability
All the simulations in this paper will be made available as markdown files on GitHub.

# Supplemental Text: Equilibria and Dynamics of Antibody Binding

## 1. The number of bound sites on a virion follows a binomial or multinomial distribution at equilibrium in absence of virus production.

*1.1 Calculating the steady state of the differential equations describing the binding of multiple types of antibodies:*

The analytical solution of virion bound sites distribution in the case of two or more types of antibodies present is derived below through solving the ordinary differential equations of the proportion of sites bound by antibody $i$ ($p_i$) at equilibrium. The following is the process of solving the ODE:

$$\frac{dp_i}{dt} = A_i K_{on_i}\left(1 - \sum_{j=1}^{m} pj\right) - K_{off_i} p_i = 0$$

$$p_i = \frac{A_i K_{on_i}}{K_{off_i}}(1 - \sum_{j=1}^{m} p_j)$$

$$\frac{A_i K_{on_i}}{K_{off_i}}(1 - \sum_{j=1}^{m} p_j) = \frac{A_i K_{on_i}}{K_{off_i}}(1 - \sum_{j=1}^{m} \frac{A_j K_{on_j}}{K_{off_j}}(1 - \sum_{j=1}^{m} p_j))$$

$$1 - \sum_{j=1}^{m} p_j = 1 - \sum_{j=1}^{m} \frac{A_j K_{on_j}}{K_{off_j}}(1 - \sum_{j=1}^{m} p_j)$$

$$1 - \sum_{j=1}^{m} p_j = 1 - (1 - \sum_{j=1}^{m} p_j) \cdot \sum_{j=1}^{m} \frac{A_j K_{on_j}}{K_{off_j}}$$

$$\left(1 - \sum_{j=1}^{m} p_j\right) + (1 - \sum_{j=1}^{m} p_j) \cdot \sum_{j=1}^{m} \frac{A_j K_{on_j}}{K_{off_j}} = 1$$

$$(1 - \sum_{j=1}^{m} p_j)(1 + \sum_{j=1}^{m} \frac{A_j K_{on_j}}{K_{off_j}}) = 1$$

$$1 - \sum_{j=1}^{m} p_j = \frac{1}{1 + \sum_{j=1}^{m} \frac{A_j K_{on_j}}{K_{off_j}}}$$

Since we earlier derived $p_i = \frac{A_i K_{on_i}}{K_{off_i}}(1 - \sum_{j=1}^{m} p_j)$, we can now get:

$$p_i = \frac{A_i K_{on_i}}{K_{off_i}} \cdot \frac{1}{1 + \sum_{j=1}^{m} \frac{A_j K_{on_j}}{K_{off_j}}}$$

## 2. Analytical solution for proportion of virion bound sites during infection

*2.1 Detailed calculation of the proportion of virion bound sites with a single type of antibody case*

To derive the analytical solution for proportion of bound sites during infection, we have first considered the proportion of sites bound on virions of age $a$ as $p(a) = \int_0^a K_{on}A(1 - p(t)) - K_{off}\, dt$. Then we defined the age distribution of virions as $f(a) = (r + d_v)e^{-ra}e^{-d_v a}$, where $r$ stands for virus growth rate and $dv$ stands for virus decaying rate. Given $P(a)$ and $f(a)$, we take the integral of the multiplication of both functions and derive the final proportion of virion bound sites $p^*$. The following is the detailed derivation of $p^*$.

We first considered the steps to derive the solution of *P(a)*:

$$\frac{dp}{dt} = K_{on}A(1 - p) - K_{off}p$$

$$1\,dt = \frac{1}{K_{on}A(1 - p) - K_{off}p}\,dp$$

$$\int_0^a 1\,dt = \int_0^a \frac{1}{K_{on}A(1 - p) - K_{off}p}\,dp$$

$$\int_0^a 1\,dt = \int_0^a \frac{1}{K_{on}A - (K_{on}A + K_{off})p}\,dp$$

$$a = \frac{-1}{K_{on}A + K_{off}}\ln\big(K_{on}A - (K_{on}A + K_{off})p(a)\big)$$

$$-\frac{-1}{K_{on}A+K_{off}}\ln(K_{on}A)$$

$$-(K_{on}A + K_{off})a = \ln\left(\frac{K_{on}A - (K_{on}A+K_{off})p(a)}{K_{on}A}\right)$$

$$e^{-(K_{on}A+K_{off})a} = \frac{K_{on}A - (K_{on}A+K_{off})p(a)}{K_{on}A}$$

$$e^{-(K_{on}A+K_{off})a}K_{on}A = K_{on}A - (K_{on}A + K_{off})p(a)$$

$$-e^{-(K_{on}A+K_{off})a}K_{on}A + K_{on}A = (K_{on}A + K_{off})p(a)$$

$$p(a) = \frac{-e^{-(K_{on}A+K_{off})a}K_{on}A + K_{on}A}{K_{on}A + K_{off}}$$

$$p(a) = \frac{(1 - e^{-(K_{on}A+K_{off})a})K_{on}A}{K_{on}A + K_{off}}$$

*Age distribution of virions f(a):*

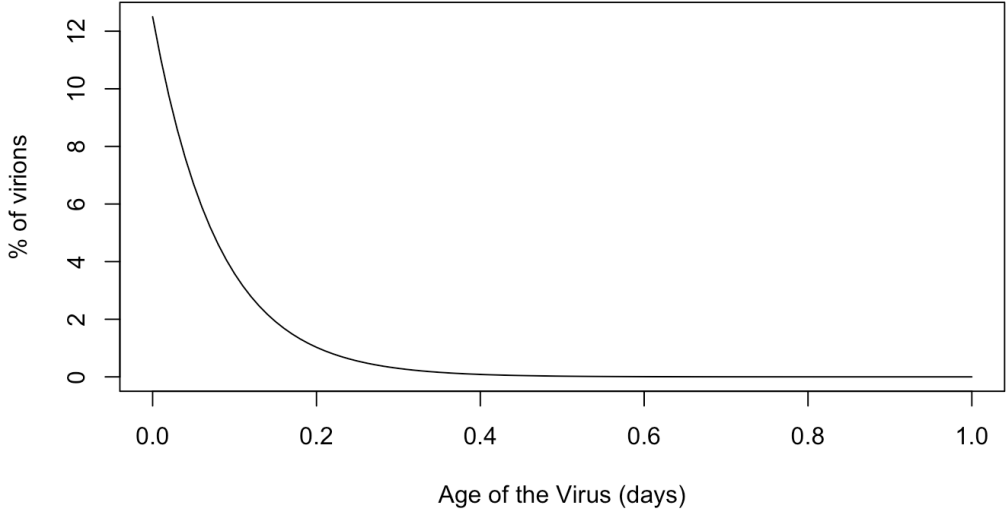

**Figure S3: Age distribution of virions under infection.** The age distribution of virions, *f(a)*, is visualized.

*Proportion of bound sites p*:

$$p^* = \int_0^\infty p(a) \cdot f(a) \, da$$

$$p^* = \int_0^\infty \frac{-e^{-(K_{on}A + K_{off})a} K_{on}A + K_{on}A}{K_{on}A + K_{off}} \cdot (r + d_v) e^{-ra} e^{-d_v a} \, da$$

$$p^* = \frac{K_{on}A(r + d_v)}{K_{on}A + K_{off}} \cdot \int_0^\infty (-e^{-(K_{on}A + K_{off})a} + 1) \cdot e^{-(r + d_v)a} da$$

$$p^* = \frac{K_{on}A(r + d_v)}{K_{on}A + K_{off}} \cdot \left( \int_0^\infty -e^{-(K_{on}A + K_{off} + r + d_v)a} da + \int_0^\infty e^{-(r + d_v)a} da \right)$$

$$p^* = \frac{K_{on}A(r + d_v)}{K_{on}A + K_{off}} \cdot \left( \frac{-1}{K_{on}A + K_{off} + r + d_v} + \frac{1}{r + d_v} \right)$$

$$p^* = \frac{K_{on}A}{K_{on}A + K_{off} + r + d_v}$$

*2.2 Detailed proof of the proportion of bound sites with multiple types of antibodies during infection*

To prove that the proportion of bound sites by antibody type $i$ is equal to $\frac{K_{on_i}A_i}{K_{off_i}+d_v+r} \cdot \frac{1}{1+\sum_{j=1}^{m}\frac{K_{on_j}A_j}{K_{off_j}+d_v+r}}$, we first defined the following parameters:

$$p_i^* = \text{proportion of sites bound by antibody type } i$$

$$= \frac{Kon_iA_i}{K_{off_i}+d_v+r} \cdot \frac{1}{1+\sum_{j=1}^{m}\frac{K_{on_j}A_j}{K_{off_j}+d_v+r}}$$

$$n = \text{total number of avaliable sites on a virion}$$

$$V = \text{number of virions present}$$

$$S_i = \text{total sites bound by antibody type } i \text{ within the virus population}$$

$$= n \cdot V \cdot p_i^*$$

Thus, the change in total sites bound by antibody type $i$ over time is the following:

$$\frac{dS_i}{dt} = K_{on_i}A_i\left(n \cdot V - \sum_{j=1}^{m}S_j\right) - K_{off_i}S_i - d_vS_i$$

$$= K_{on_i}A_i \cdot n \cdot V\left(1 - \sum_{j=1}^{m}p_j^*\right) - K_{off_i} \cdot n \cdot V \cdot p_i^* - d_v \cdot n \cdot V \cdot p_i^*$$

$$= n \cdot V\left[K_{on_i} \cdot A_i \cdot \left(1 - \sum_{j=1}^{m}p_j^*\right) - K_{off_i}p_i^* - d_vp_i^*\right]$$

$$= n \cdot V\left[\frac{K_{on_i}A_i}{1+\sum_{j=1}^{m}\frac{K_{on_j}A_j}{K_{off_j}+d_v+r}} - \left(K_{off_i}+d_v\right)\frac{K_{on_i}A_i}{K_{off_i}+d_v+r}\right.$$

$$\left. \cdot \frac{1}{1+\sum_{j=1}^{m}\frac{K_{on_j}A_j}{K_{off_j}+d_v+r}}\right]$$

$$= n \cdot V\left[\frac{K_{on_i}A_i - \left(K_{off_i}+d_v\right)\frac{K_{on_i}A_i}{K_{off_i}+d_v+r}}{1+\sum_{j=1}^{m}\frac{K_{on_j}A_j}{K_{off_j}+d_v+r}}\right]$$

$$= n \cdot V\left[\frac{K_{on_i}A_i\left(1 - \frac{K_{off_i}+d_v}{K_{off_i}+d_v+r}\right)}{1+\sum_{j=1}^{m}\frac{K_{on_j}A_j}{K_{off_j}+d_v+r}}\right]$$

$$= n \cdot V \left[ \frac{K_{on_i} A_i \left( \frac{K_{off_i} + d_v + r - K_{off_i} - d_v}{K_{off_i} + d_v + r} \right)}{1 + \sum_{j=1}^{m} \frac{K_{on_j} A_j}{K_{off_j} + d_v + r}} \right]$$

$$= r \cdot n \cdot V \frac{\frac{K_{on_i} A_i}{K_{off_i} + d_v + r}}{1 + \sum_{j=1}^{m} \frac{K_{on_j} A_j}{K_{off_j} + d_v + r}}$$

$$= r \cdot n \cdot V \cdot p_i^*$$

$$= r \cdot S_i$$

The infection is growing at rate $r$, and if everything else (the $S_i$ state variables) is also increasing at this rate, then they have fallen into an eigenvector. Thus, it proves that the proportion of bound sites by antibody type $i$ is approaching to $\frac{K_{on_i} A_i}{K_{off_i} + d_v + r} \cdot \frac{1}{1 + \sum_{j=0}^{m} \frac{K_{on_j} A_j}{K_{off_j} + d_v + r}}$ as the system falls into the eigenvector.

# 3. The virus infectivity decreases in a convex manner as proportion of bound sites increases

### 3.1 Schematic of the WNV experiment

To derive the function of virus infectivity and proportion of bound sites, we use the empirical data from the work of Pierson et.al. Detailed experiment procedure is described in the main text, and the schematic of the experiment is shown in the following Figure S4.

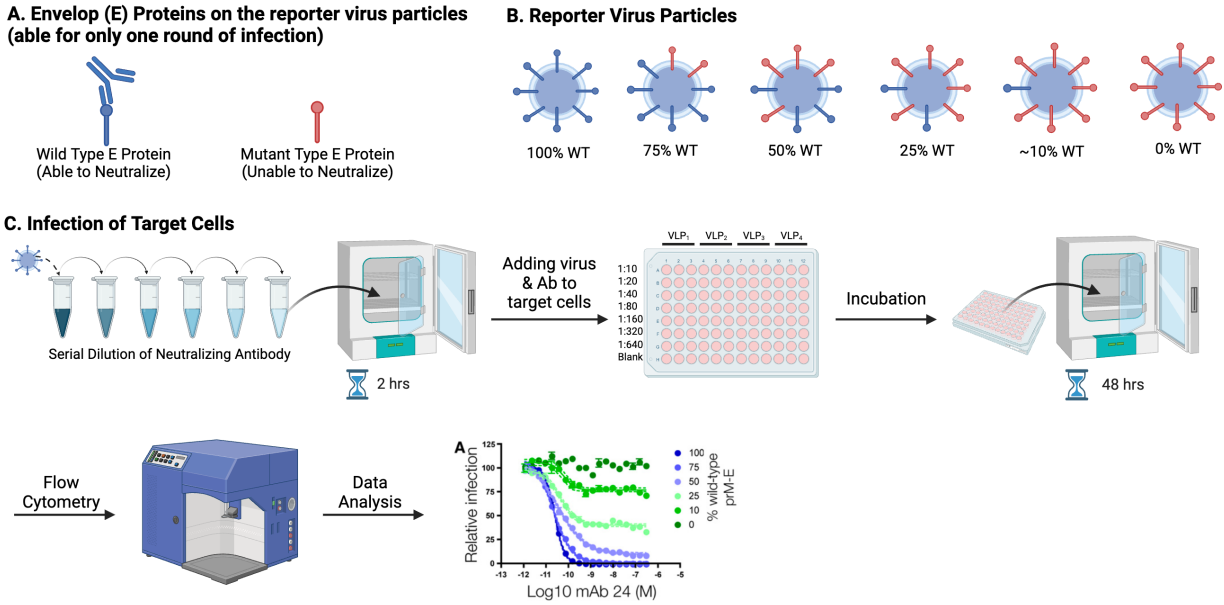

**Figure S4: Schematic of the WNV experiment.** We utilized Biorender to visually represent the experiment procedure, drawing upon the material and methods outlined in Pierson et al.'s paper, from which we extracted empirical data for our study.

### 3.2 Justification of relative infection is proportional to relative infectivity

#### 3.2.1 Mathematical proof

We first define that relative infection is the ratio between the number of infected cells when the antibody is present or not. Relative infectivity is the ratio between the virus infection rate when the antibody is present or not. Thus, to show that the two are approximately proportional, we use an ODE model that depicts basic virus, infected cell, and target cell population dynamics (Model S1). Now, relative infection at time $\tau$ is $\frac{I(\tau)_{Ab}}{I(\tau)_{max}}$, whereas relative infectivity is $\frac{\beta_{Ab}}{\beta_{max}}$.

$$\frac{dU}{dt} = -\beta UV$$
$$\frac{dI}{dt} = \beta UV \qquad (Model\ S1)$$

$$\frac{dV}{dt} = -d_v V$$

We then solve the for both U and V at $t = \tau$:

$$V(\tau) = V(0)e^{-d_v \tau}$$

$$\ln(U(\tau)) = \frac{\beta V(0)}{d_v} e^{-d_v \tau} = \beta \cdot \frac{V(0)}{d_v} e^{-d_v \tau}$$

Thus, relative infection $\frac{\beta_{Ab}}{\beta_{max}}$ can be expressed in terms of the following:

$$\frac{\beta_{Ab}}{\beta_{max}} = \frac{\ln(U(\tau)_{Ab})}{\ln(U(\tau)_{max})} = \frac{\ln(U(0) - I(\tau)_{Ab})}{\ln(U(0) - I(\tau)_{max})}$$

Although Pierson et al do not always report the number of infected cells, in those cases where they are reported, the infected cells proportions are always lower than 25% and typically much lower. When the number of infected cells $I$ is relatively low compared the number of target cells $U$, we can use the first order Taylor series expansion:

$$\frac{\ln(U(0) - I(\tau)_{Ab})}{\ln(U(0) - I(\tau)_{max})} \approx \frac{I(\tau)_{Ab}}{I(\tau)_{max}}$$

Thus, relative infectivity is approximately proportional to relative infection:

$$\frac{\beta_{Ab}}{\beta_{max}} = \frac{\ln(U(\tau)_{Ab})}{\ln(U(\tau)_{max})} = \frac{\ln(U(0) - I(\tau)_{Ab})}{\ln(U(0) - I(\tau)_{max})} \approx \frac{I(\tau)_{Ab}}{I(\tau)_{max}}$$

## 4. Alternative $\beta(i)$ function

*4.1 Alternative curve shapes for the infectivity function β(i)*

Here we introduce two alternative curve shapes for the virus infectivity function: linear and concave (Figure S5). Subsequently, we will assess whether the utilization of these virus infectivity functions alters the observed patterns.

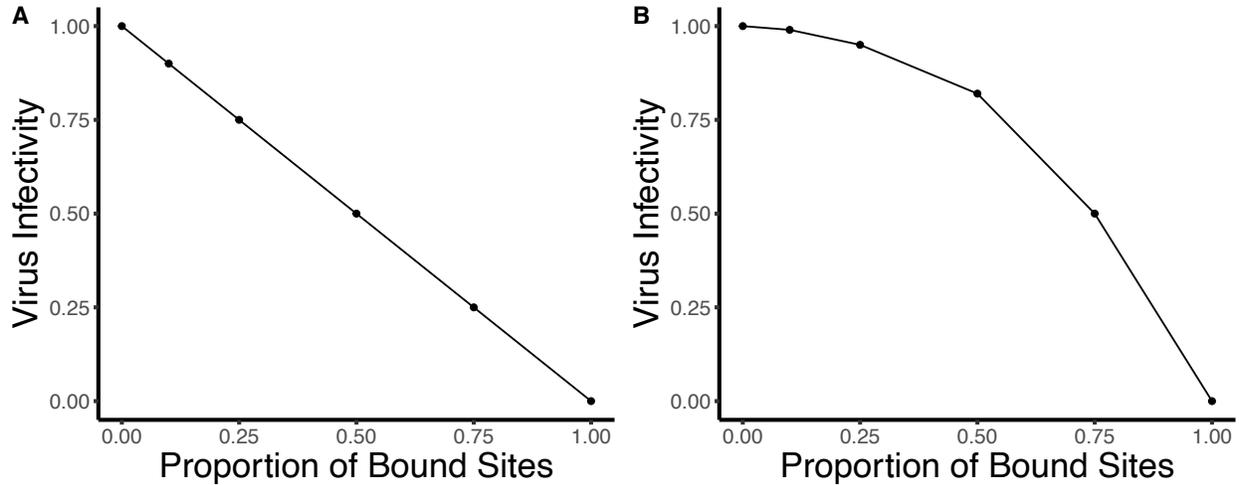

**Figure S4. Two alternative curve shapes of the virus infectivity curve. (A)** The proportion of bound sites and virus infectivity has a linear relationship. **(B)** Virus infectivity decreases in a concave manner when the proportion of bound sites increases.

*4.2 Results from the full model and the magic model using alternative infectivity curves*

As expected, the linear $\beta(i)$ curve results in a larger SLA compared to the convex curve, and when the virus infectivity curve is concave, the SLA is even larger. Other patterns are similar to the main text. The neutralization potency of the antibody as well as the effect of $K_{off}$ are reduced in the full model compared to the magic model, which is meant to reflect neutralization assays especially those with longer period of pre-incubation of antibody and virus before target cells are introduced. Similar to the main text, the amount of antibody needed to reduce viral growth by 50% is substantially smaller than the SLA (20% ~ 27%) suggesting substantial protection from even low levels of antibody. The alternative infectivity curves do, however, suggest an explanation for the discrepancy between IAA2 and SLA. Using the linear $\beta(i)$ curve the fold effect of $K_{off}$ on IAA2 and $SLA^{-1}$ are identical (Table S1), whereas with concave $\beta(i)$ the fold effect on $SLA^{-1}$ is greater (Table S2) which is the opposite to the pattern seen in the main text with convex $\beta(i)$.

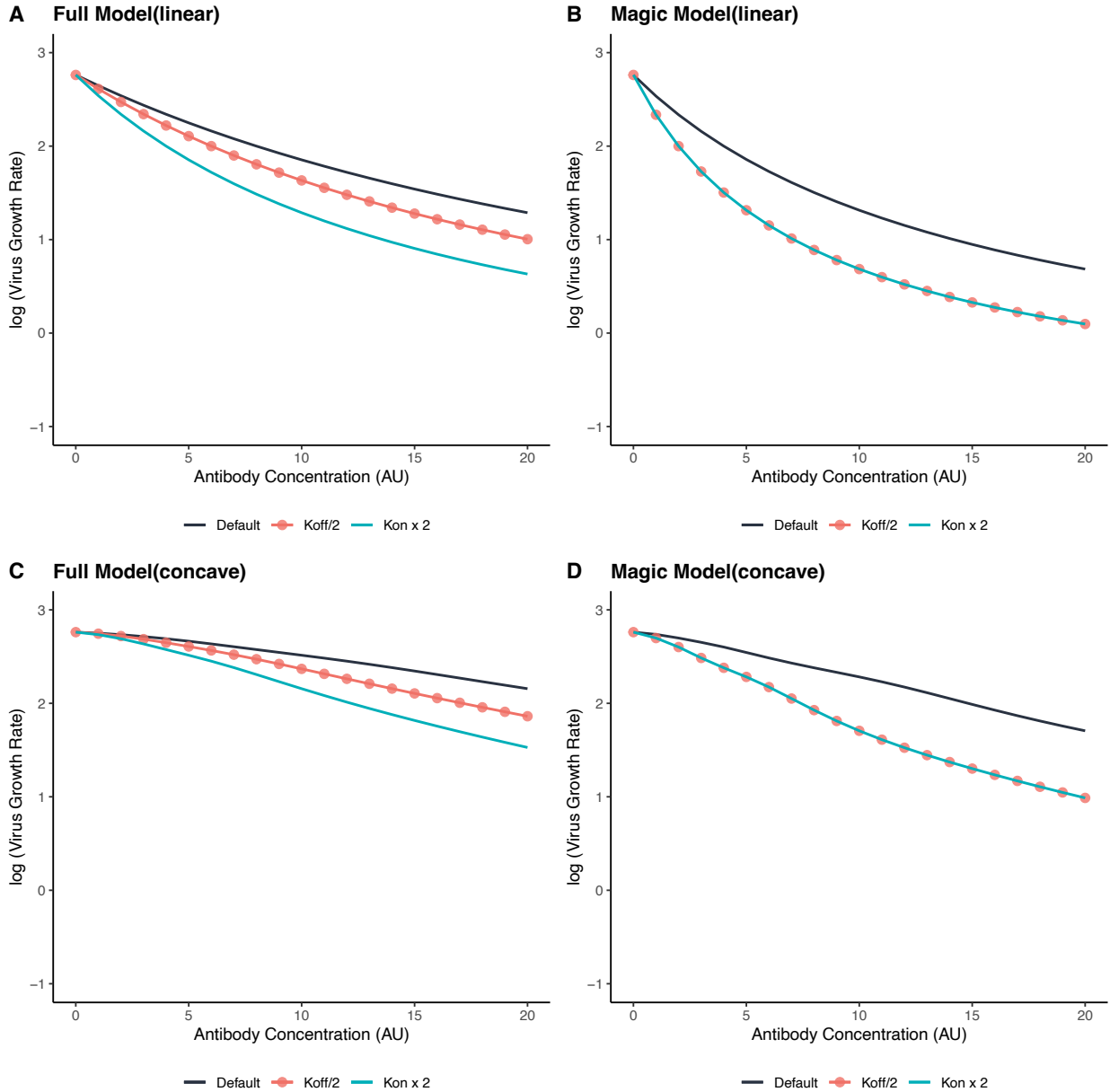

**Figure S5. Comparison between the magic model and the full model using the two alternative virus infectivity curve shapes.** Simulations regard to the change in virus growth rate (in natural log scale) as antibody concentration increases using **(A)** the magic model with linear virus infectivity curve, **(B)** full model with linear virus infectivity curve, **(C)** the magic model with concave virus infectivity curve, **(D)** the full model with concave virus infectivity curve. The black line represents the simulation conducted using default parameters. The red line represents the simulation conducted with the parameter value of $K_{off}$ set to be half the default value. The green line represents the simulation conducted with the parameter value of $K_{on}$ set to be double the default value.

**Table S1: Linear Infectivity Curve**

| Value of $K_{off}$ (day$^{-1}$) | SLA$^{-1}$ Calculated Using Full model | SLA$^{-1}$ Calculated Using Magic Model | IAA2 ($\frac{K_{on}}{K_{off}+d_v}$) | Affinity ($\frac{K_{on}}{K_{off}}$) |
|---|---|---|---|---|
| 6 | 0.016 (+1.38) | 0.044 (+2.0) | 0.0625 (+1.38) | 0.17 (+2.0) |
| 12 (default) | 0.012 | 0.022 | 0.045 | 0.083 |
| 24 | 0.0077 (-1.55) | 0.011 (-2.0) | 0.029 (-1.55) | 0.042 (-2.0) |

SLA is the minimum antibody concentration for sterilizing immunity. IAA2 is the simplified infection analog of affinity. The numbers in the parenthesis are the fold change relative to the default value with + indicated fold increase and - fold decrease. The units for SLA$^{-1}$ and affinity are AU$^{-1}$.

**Table S2: Concave Infectivity Curve**

| Value of $K_{off}$ (day$^{-1}$) | SLA$^{-1}$ Calculated Using Full model | SLA$^{-1}$ Calculated Using Magic Model | IAA2 ($\frac{K_{on}}{K_{off}+d_v}$) | Affinity ($\frac{K_{on}}{K_{off}}$) |
|---|---|---|---|---|
| 6 | 0.00087 (+1.44)$^2$ | 0.019 (+2.0) | 0.0625 (+1.38) | 0.17 (+2.0) |
| 12 (default) | 0.0069 | 0.0097 | 0.045 | 0.083 |
| 24 | 0.0037 (-1.62) | 0.0048 (-2.0) | 0.029 (-1.55) | 0.042 (-2.0) |